\documentclass[]{spie}  

\usepackage{amsmath,amsfonts,amssymb}
\usepackage{graphicx}
\usepackage[colorlinks=true, allcolors=blue]{hyperref}
\usepackage{float}

\title{Focus diverse phase retrieval testbed development of continuous wavefront sensing for space telescope applications}

\author[a]{Hyukmo Kang}
\author[b]{Kyle Van Gorkom}
\author[b]{Jess Johnson}
\author[b]{Ole Singelstad}
\author[b]{Aaron Goldtooth}
\author[a,b]{Daewook Kim}
\author[b]{Ewan S.Douglas}

\affil[a]{James C. Wyant College of Optical Sciences, University of Arizona, 1630 E. University Blvd., Tucson, AZ 85721 USA}
\affil[b]{Department of Astronomy, University of Arizona, 933 N. Cherry Ave., Tucson, AZ, 85721 USA}

\authorinfo{Further author information: (Send correspondence to Hyukmo Kang)\\Hyukmo Kang: E-mail: hkang04@arizona.edu\\  Daewook Kim: E-mail: dkim@optics.arizona.edu \\ Ewan S.Douglas: E-mail: douglase@arizona.edu}

\begin{document} 
\maketitle

\begin{abstract}
Continuous wavefront sensing on future space telescopes allows relaxation of stability requirements while still allowing on-orbit diffraction-limited optical performance. We consider the suitability of phase retrieval to continuously reconstruct the phase of a wavefront from on-orbit irradiance measurements or point spread function (PSF) images. As phase retrieval algorithms do not require reference optics or complicated calibrations, it is a preferable technique for space observatories, such as the Hubble Space Telescope or the James Webb Space Telescope. To increase the robustness and dynamic range of the phase retrieval algorithm, multiple PSF images with known amount of defocus can be utilized. In this study, we describe a recently constructed testbed including a 97 actuator deformable mirror, changeable entrance pupil stops, and a light source. The aligned system wavefront error is below $\approx 30$ nm. We applied various methods to generate a known wavefront error, such as defocus and/or other aberrations, and found the accuracy and precision of the root mean squared error of the reconstructed wavefronts to be less than $\approx 10$ nm and $\approx 2$ nm, respectively. Further, we discuss the signal-to-noise ratios required for continuous dynamic wavefront sensing. We also simulate the case of spacecraft drifting and verify the performance of the phase retrieval algorithm for continuous wavefront sensing in the presence of realistic disturbances.  
\end{abstract}


\section{INTRODUCTION}



Phase retrieval is a technique that allows the recovery of phase information in an optical system from one or more intensity measurements combined with \textit{a priori} information that constrains the possible solution space. Phase retrieval does not require an additional metrology system, as it utilizes point-spread function (PSF) images at the focal plane. This feature has advantages in space telescope applications because it reduces hardware complexity and weight. Therefore, from the Hubble Space Telescope \cite{fienup1993hubble} to the James Webb Space Telescope \cite{dean2006phase}, phase-retrieval techniques are widely used in multiple space telescope missions to provide the optical status of the telescopes. 

Focus diversity phase retrieval (FDPR) makes use of multiple defocused PSFs to break the degeneracy in the pupil-plane phase solution, to increase robustness against noise and other sources of error\cite{fienup2013phase}, and to increase sensitivity to particular spatial frequency content\cite{dean2003diversity}.

The particular variant of FDPR employed in the testbed follows the reverse-mode algorithmic differentiation (RMAD) approach first applied to phase retrieval by Jurling\cite{jurling2014applications}. A numerical forward diffraction model propagates wavefront error in the pupil-plane to intensity in the focal-plane or in a defocused plane. An error metric is defined to quantify the error in the model fit to the measured intensity and minimized with a non-linear solver. Numerical gradients are calculated efficiently via RMAD and then passed to the solver. The model is parameterized by a set of modal coefficients, or the pixel-by-pixel phase values, but in principle any number of parameters that describe the optical setup or unknowns can be fit.

\section{Testbed setup}
\label{sec:Testbed setup}

\begin{figure}[H]
\centering\includegraphics[width=10cm]{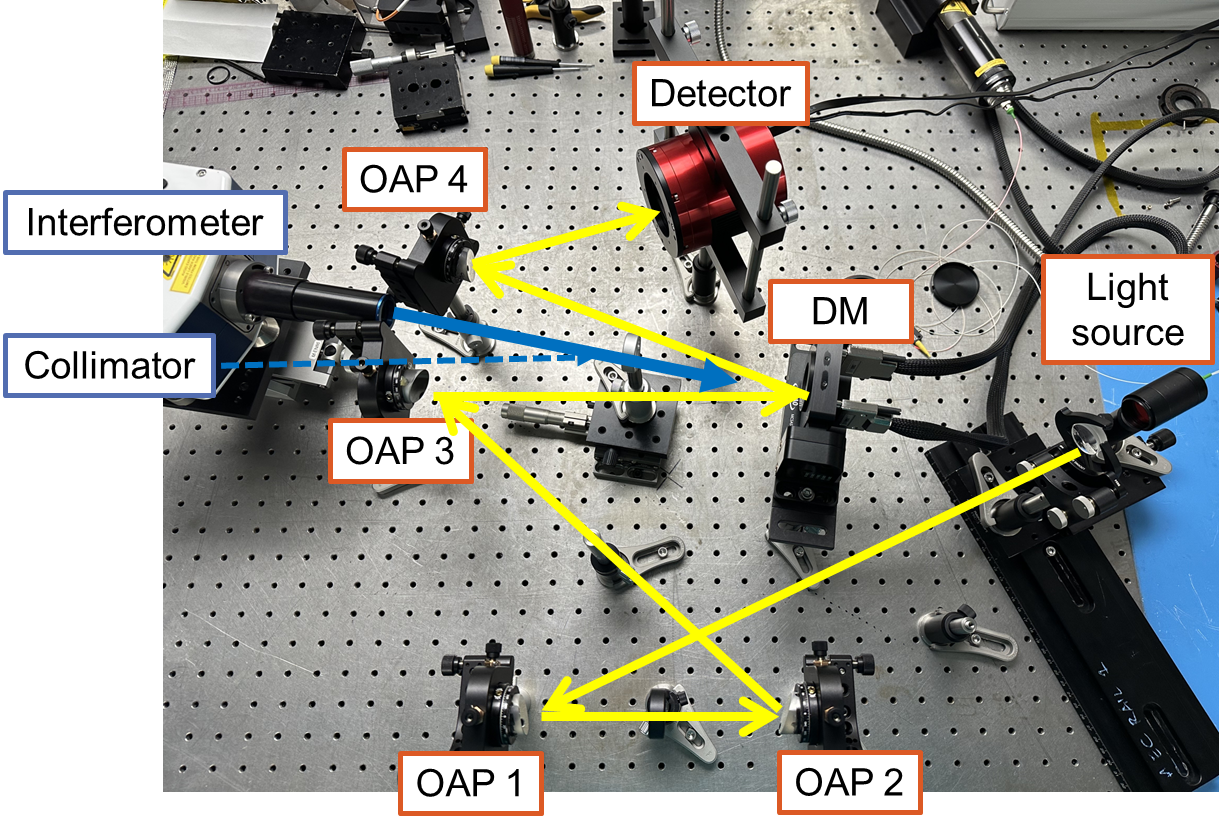}
\caption{Layout of the testbed. Yellow arrows represent the path of the light from the HeNe laser source. Blue arrows represent the optical path of the interferometer.}
\label{fig:testbed}
\end{figure}

We have built a testbed to implement our FDPR algorithm (Fig.~\ref{fig:testbed}). We made a point source using a 4-f system with a HeNe laser fiber and an achromatic lens. This setup enables us to switch the light source to a broadband light source, which is a part of our future work. The light from the HeNe is collimated by the first off-the-shelf off-axis parabolic mirror (OAP), OAP 1, which has a focal length of 304.8~mm and an off-axis distance (OAD) of 119.4~mm. The collimated beam passes through the pupil mask and is reimaged by OAP 2, which has a focal length of 152.4~mm and an OAD of 101.6~mm. The beam is now recollimated with OAP 3, which has the same specifications as OAP 2, and then hits a deformable mirror (DM), which has a 13.5~mm aperture and 97 actuators across the aperture. The position of the DM is conjugate to the pupil plane. Finally, the reflected beam from the DM hits OAP 4, which also has the same specifications as OAP 2 and 3, and then forms an image at the detector. We also used an interferometer with a collimating lens aimed at the DM's surface to monitor the DM's surface shape during the FDPR algorithm tests. The testbed's components are summarized in  Table~\ref{tab:compo}.

\begin{table}[H]
\caption{List of Testbed Components}
\label{tab:compo}
\begin{center}
\begin{tabular}{ll}
\hline
Components     & Notes                                                 \\ \hline
Light   source & 632.8~nm                                            \\
OAP   1        & Focal length:   304.8~mm (12"), OAD: 119.4 mm (4.7")            \\
OAP   2, 3, 4  & Focal length:   152.4~mm (6"), OAD: 101.6 mm (4")               \\
DM             & ALPAO   97-15 (97 actuators, pupil diameter: 13.5~mm) \\
Camera       & ZWO ASI6200 (Detector: SONY IMX455, 3.76~µm pixel size)       \\ 
Interferometer & 4D PhaseCam 6000 \\ \hline                          
\end{tabular}
\end{center}
\end{table}

\begin{figure}[H]
\centering\includegraphics[width=11cm]{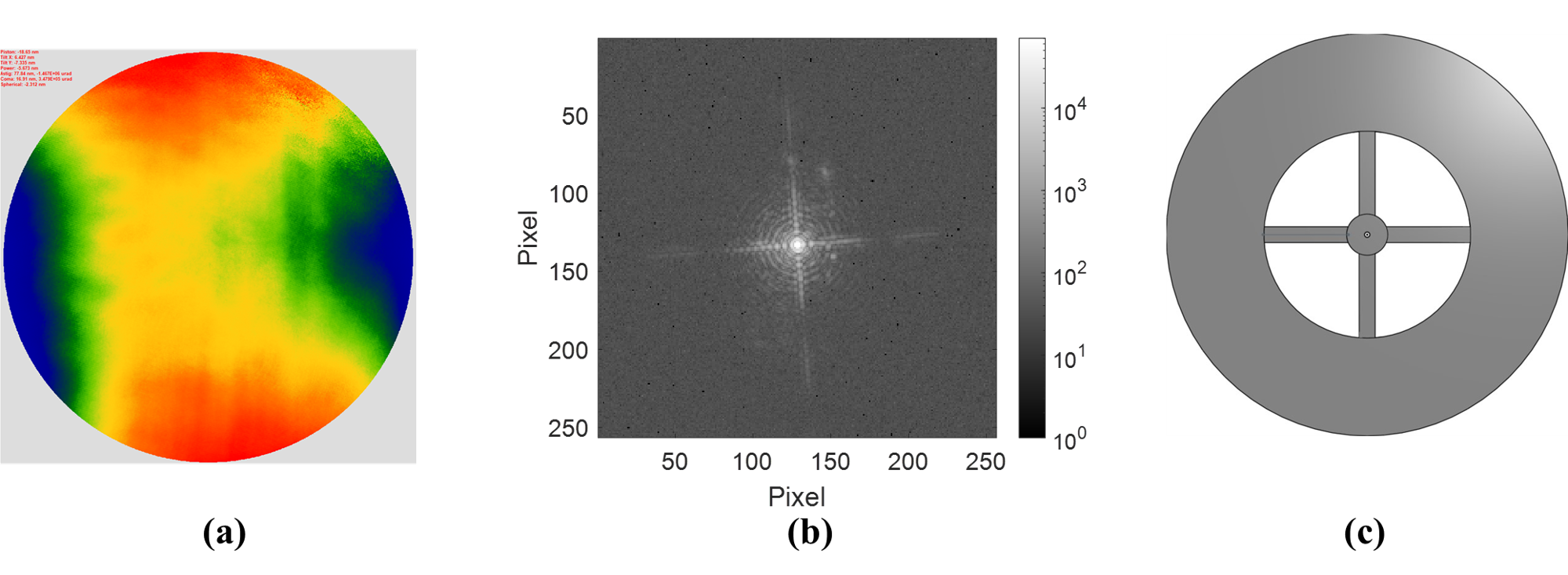}
\caption{(a) Wavefront error map resulting from the initial alignment of the testbed. (b) Point spread function (PSF) after fine adjustment of the deformable mirror. (c) Design of the pupil mask with central obscuration and spider mount.}
\label{fig:initial_alignment}
\end{figure}

The results of aligning the testbed are shown in Fig.~\ref{fig:initial_alignment}a. The root-mean-square (RMS) wavefront error (WFE) after alignment is 29~nm (Fig.~\ref{fig:initial_alignment}a). After adjusting the DM, we achieved a PSF (Fig.~\ref{fig:initial_alignment}b) which we will use as a reference PSF in the following experiments. Also, we used a pupil mask with a diameter of 13~mm and a 10\% central obscuration with four 1~mm thick struts (Fig.~\ref{fig:initial_alignment}c) so that the cross-like diffraction pattern of a secondary's spider mount appears in the PSF. 

\section{Testbed results}
\label{sec:Testbed results}

\begin{figure}[H]
\centering\includegraphics[width=12cm]{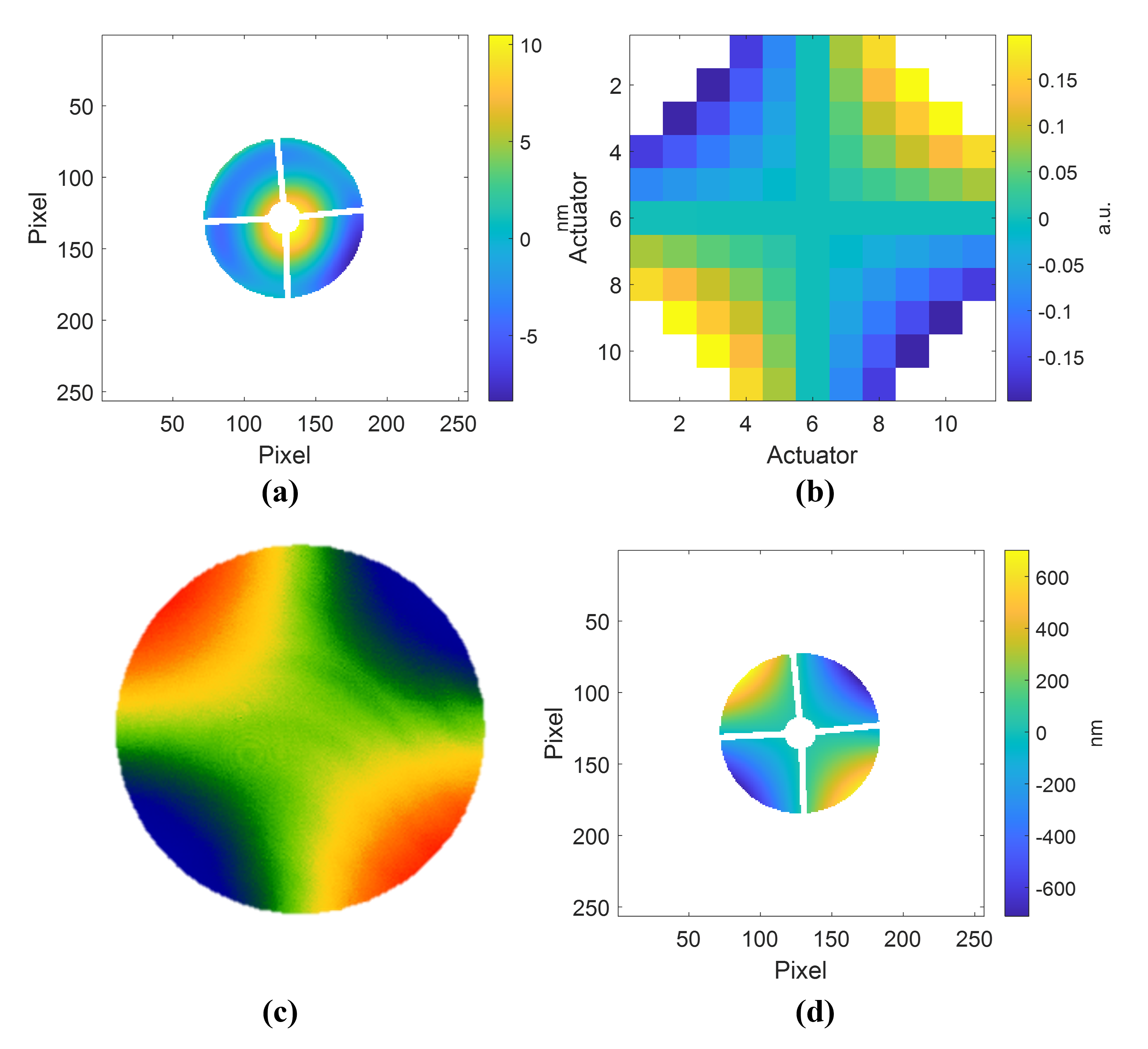}
\caption{Initial focus diversity phase retrieval results. (a) Processed wavefront error map after testbed alignment and deformable mirror adjustments. (b) DM input shape for the accuracy test. (c) WFE map from the interferometer; root-mean square WFE is 148.7~nm. (d) WFE map from 10 repetitions of FDPR measurements. RMS WFE is 150.22~nm with $\sigma=0.69$~nm.}
\label{fig:initial_fdpr_result}
\end{figure}

To validate the performance of the FDPR algorithm, we first evaluated the repeatability of experiments on the testbed. For testing focus diversity, we used the DM to generate -0.075~$\lambda$, -0.03~$\lambda$ and 0.075~$\lambda$ defocused PSFs from the reference PSF (Fig.~\ref{fig:initial_alignment}b). Fig.~\ref{fig:initial_fdpr_result}a shows the reconstructed WFE map produced by FDPR. After repeating measurements ten times, the resulting RMS WFE was 3.74~nm with $\sigma=0.14$~nm. This WFE map will be used as a reference map for the following measurements.

The accuracy of the FDPR algorithm is evaluated by comparing its results with those of the interferometer. We added astigmatism aberration to the beam by changing the DM shape (Fig.~\ref{fig:initial_fdpr_result}b and measured the WFE with the interferometer (Fig.~\ref{fig:initial_fdpr_result}c). The RMS WFE measurement from the interferometer was 148.7~nm, and the Zernike coefficient term corresponding to astigmatism (Z5) was 148.24~nm. The FDPR algorithm showed 150.22~nm of RMS WFE with $\sigma=0.69$~nm and 149.98~nm of Z5. This is a difference of  $\approx $ 1\%.

\begin{figure}[H]
\centering\includegraphics[width=12cm]{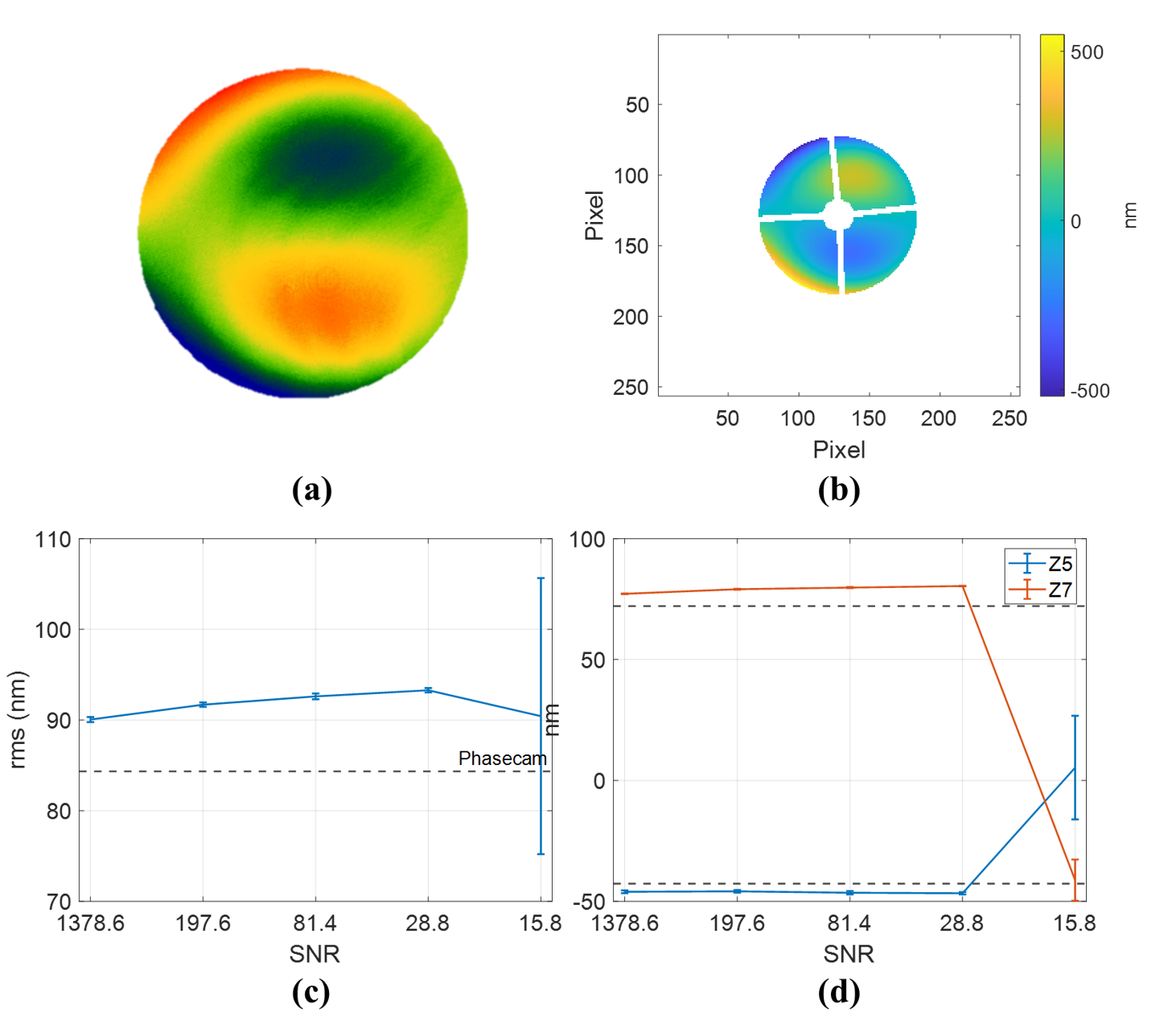}
\caption{Results of the signal-to-noise ratio tests. (a) The measured WFE map from the interferometer. (b) The retrieved WFE map from the FDPR algorithm when SNR is set to 1378.6. (c) SNR vs. RMS. Error bars in the figure represents standard deviations. (d) SNR vs. Zernike coefficients (Z5: Astigmatism, Z7: Coma).}
\label{fig:snr_trends}
\end{figure}

We also tested how variations in the signal-to-noise ratio (SNR) affect the accuracy of the FDPR algorithm. In this experiment, we calculated the SNR by dividing the maximum value of the PSF by its mean value. We introduced astigmatism and coma aberration (Fig.\ref{fig:snr_trends}a). The RMS WFE measured with the interferometer was 84.34~nm and the corresponding Zernike coefficient terms were -42.66~nm astigmatism (Z5), and 72.12~nm coma (Z7). Fig.~\ref{fig:snr_trends}b shows the reconstructed WFE produced by FDPR, which has 90.06~nm RMS ($\sigma$ = 0.22~nm), a Z5 measurement of -46.02~nm ($\sigma$ = 0.57~nm), and a Z7 measurement of 77.19~nm ($\sigma$ = 0.14~nm) when the SNR is 1378.6. 

We then varied the SNR five times and measured the WFE in a similar way for each different SNR value. We varied the SNR by reducing the intensity of the light source using a fiber-attached optical variable attenuator. Figs.~\ref{fig:snr_trends}c and~\ref{fig:snr_trends}d show the results of the FDPR algorithm. What this shows is that When the SNR value is higher than 28.8, both the RMS and Zernike coefficients are within 3 nm, even though there is an offset between the interferometer and algorithm results. On the other hand, when the SNR drops below 15.8, the accuracy of the FDPR algorithm begins to fail, and neither the RMS or Zernike coefficients are accurate.

\begin{figure}[H]
\centering\includegraphics[width=9cm]{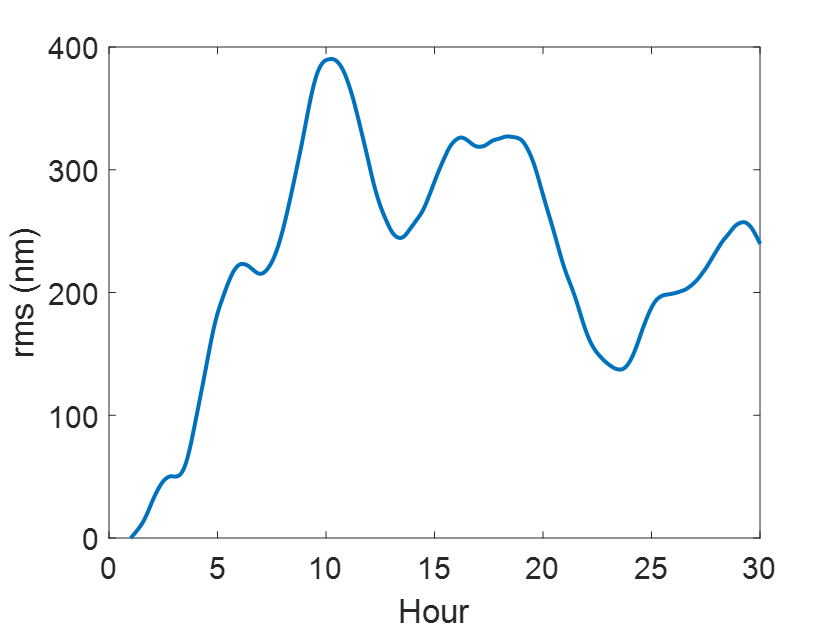}
\caption{Synthetic data for the telescope drifting case study.\cite{douglas2023}}
\label{fig:raw_simulation}
\end{figure}

Finally, we tested applying the FDPR algorithm continuously to correct aberration changing over time in a hypothetical realization of a statistical telescope drifting scenario. Fig.~\ref{fig:raw_simulation} shows the simulated varying RMS values during 30 hours of orbital operation\cite{douglas2023}. Assuming that gravitational and thermal deformation would mainly affect low-order shapes, we imposed defocus, astigmatism, coma, trefoil, and spherical aberrations only, and ignored higher order aberrations. Without this type of correction, the in-flight performance of a telescope slowly diverges and is unable to achieve diffraction-limited quality. 

\begin{figure}[H]
\centering\includegraphics[width=12cm]{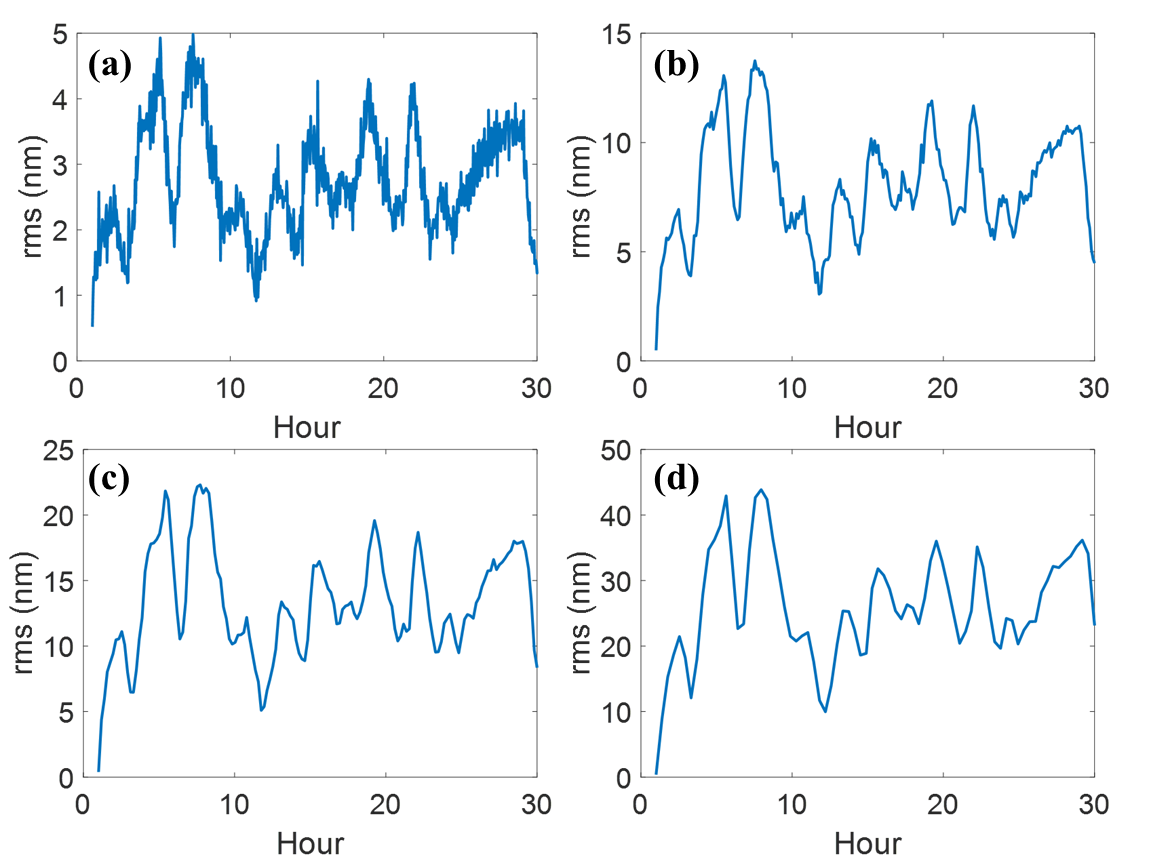}
\caption{Residual RMS WFE for different temporal lag cases. The temporal gap between measurement and correction is (a) 140 seconds (b) 520 seconds (c) 700 seconds (d) 1400 seconds. The residual RMS is smaller in shorter temporal gap cases.}
\label{fig:drift_result}
\end{figure}

Fig.~\ref{fig:drift_result} represents the RMS values of the residual WFE of continuous wavefront detection and corrections under multiple correction scenarios. Even though we are aiming for real time correction, there will be a temporal lag between the data acquisition for the FDPR algorithm and the application of physical correction. We evaluated the performance of the FDPR algorithm while increasing the temporal lag from 140 seconds (Fig.~\ref{fig:drift_result}a) to 1400 seconds (Fig.~\ref{fig:drift_result}d). As expected, the residual RMS WFE is restrained more efficiently with a shorter temporal gap. 

\section{Summary and Discussion}
\label{sec:Discussion}

In this study, we built a testbed to evaluate the performance of the FDPR algorithm in different situations. The demonstrated accuracy of the reconstructed RMS WFE is less than 10~nm with less than 2~nm of precision. We determined that the minimum SNR required for the algorithm on the testbed is larger 15.8. Also, we tested the performance of the FDPR algorithm for continuous wavefront sensing in a simulated telescope drifting case with various temporal gaps between measurement and correction. The results from the required SNR and the drifting scenario will help us to generate a list of guide stars of appropriate magnitude that we can use in orbit for continuous wavefront sensing.

As discussed in this paper, having an SNR over a particular threshold is essential to effectively using the phase-retrieval technique. However, increasing exposure time necessarily introduces longer temporal gaps between measurements and their associated corrections. Taking additional defocused images would increase the temporal gap as well as system complexity. This leads to the next goal of our continuing work, which is finding a strategy to ensure that we can achieve the required SNR while simultaneously shortening image acquisition time.

\acknowledgments 
 
Portions of this research were supported by funding from the Technology Research Initiative Fund (TRIF) of the Arizona Board of Regents and by generous anonymous philanthropic donations to the Steward Observatory of the College of Science at the University of Arizona. 

\bibliography{report} 
\bibliographystyle{spiebib} 

\end{document}